\documentclass[preprintnumbers,amsmath,amssymb,psfig,preprint]{revtex4}
\usepackage{graphicx}

\begin{document}

\title{Some Entanglement Features of Highly Entangled Multiqubit States}

\author{A. BORRAS, M. CASAS}

\address{Departament de F\'{\i}sica and IFISC-CSIC, Universitat de les Illes Balears\\
07122 Palma de Mallorca, Spain \\
toni.borras@uib.es, montse.casas@uib.es}

\author{A.R. PLASTINO$^{\dag}$}

\address{Physics Department, University of Pretoria\\
Pretoria 0002, South Africa \\
arplastino@maple.up.ac.za}

\author{A. PLASTINO}

\address{$^{\dag}$National University La Plata-CONICET\\
C.C. 727, 1900 La Plata, Argentina \\
plastino@fisica.unlp.edu.ar}

\vspace{5cm}

\begin{abstract}
We explore some basic entanglement features of multiqubit
systems that are relevant for the development of
algorithms for searching highly entangled states.
In particular, we compare the behaviours of multiqubit
entanglement measures based (i) on the von Neumann entropy of marginal
density matrices and (ii) on the linear entropy of those matrices.
\end{abstract}

\keywords{Multiqubit Entanglement, Highly Entangled States.}

\maketitle

\section{Introduction}

The study of quantum entanglement is contributing
both to the elucidation of the foundations of
quantum mechanics and  to the birth
of new, revolutionary technologies \cite{PV07,BZ06,NC00}.
A considerable amount of research has recently been devoted
to the study of multiqubit entanglement measures defined as
the sum of bipartite entanglement measures over all
(or an appropriate family of) the possible bi-partitions of
the full system \cite{WH05,CMB04,AM06,CHDB05}.
The aim of the present contribution is to explore some
basic properties of highly entangled multiqubit states,
and also of the ``entanglement landscape" in their
neighbourhoods.  We compare the behaviours of two
entanglement measures for multiqubit pure states,
one based on the von Neumann entropy of marginal density matrices
and the other based upon the linear entropy of those matrices.
We also compare the performances of two searching algorithms for
highly entangled states, based on different families of
bi-partitions of the multiqubit system.

\section{Multiqubit Entanglement}

 The genuine multipartite entanglement $E$ of a $N$-qubit state can be
expressed as

\begin{eqnarray}
E &=& \frac{1}{[N/2]} \sum_{m=1}^{[N/2]} E^{(m)}, \\
E^{(m)} &=& \frac{1}{N_{bipart}^m}
\sum_{i=1}^{N_{bipart}^m} E(i). \label{Entsub}
\end{eqnarray}

Here, $E(i)$ stands for the entanglement associated with one,
single bi-partition of the $N$-qubits system. The quantity
$E^{(m)}$ gives the average entanglement between subsets
of $m$ qubits and the remaining $N-m$ qubits constituting the system.
The average is performed over the $N_{bipart}^{(m)}$ nonequivalent
ways to do such bi-partitions, which are given by

\begin{eqnarray}
N_{bipart}^{m} &=& \binom{N}{n}\textrm{     if }n \neq N/2,\\
N_{bipart}^{N/2} &=& \frac{1}{2} \binom{N}{N/2}\textrm{     if }n = N/2.
\end{eqnarray}

\noindent Different $E^{(m)}$ represent different entanglement properties
of the state. While $E^{(1)}$ can attain its maximum value for a given
state, $E^{(2)}$ can be arbitrarily low for such state. This is why all
these entanglement measures must be computed to capture all the
entanglement properties of the state. The global multiqubit entanglement
is given by the average of the $[N/2]$ different $E^{(m)}$ for any
state $|\Psi\rangle$.

We will use two types on entanglement measures, $E_L$ and $E_{vN}$,
respectively based on two different measures for the
mixedness of the marginal density matrices $\rho_i$ associated
with the bi-partitions: (i) the linear entropy $S_L = \frac{2^m}{2^m-1}(1-Tr[\rho_i^2])$,
and (ii) the von Neumann entropy $S_{vN}= -Tr [\rho_i log \rho_i]$.
If one uses the linear entropy $S_L$, $E^{(1)}_L$ turns out to be the
well known Meyer-Wallach multipartite entanglement measure \cite{MW02}
that Brennen \cite{Brennen03} showed to coincide with the average of
all the
single-qubit linear entropies. This measure was later generalized
by Scott \cite{Scott04} to the case where all possible bi-partitions
of the system where considered.

The entanglement measure given in Eq. (1) is maximized by a state
which has all its reduced density matrices maximally mixed. Although
it is easy to verify that in the 3 qubit case $|GHZ \rangle$
complies with this requirement, the situation becomes much more
complicated when systems with four or more qubits are considered.
 Higuchi and Sudbery proved that there is no 4 qubit state whose
 two qubit reduced density matrices are all maximally mixed. They
 conjectured that the 4 qubit state exhibiting the higher entanglement
 is

\begin{equation}
|HS\rangle \, = \, \frac{1}{\sqrt{6}} \Bigl[ |1100\rangle+
|0011\rangle + \omega \Bigl(|1001\rangle + |0110\rangle \Bigr) +
\omega^2 \Bigl(|1010\rangle + |0101\rangle\Bigr) \Bigr], \label{HS}
\end{equation}

\begin{figure}
\begin{center}
\vspace{0.5cm}
\includegraphics[]{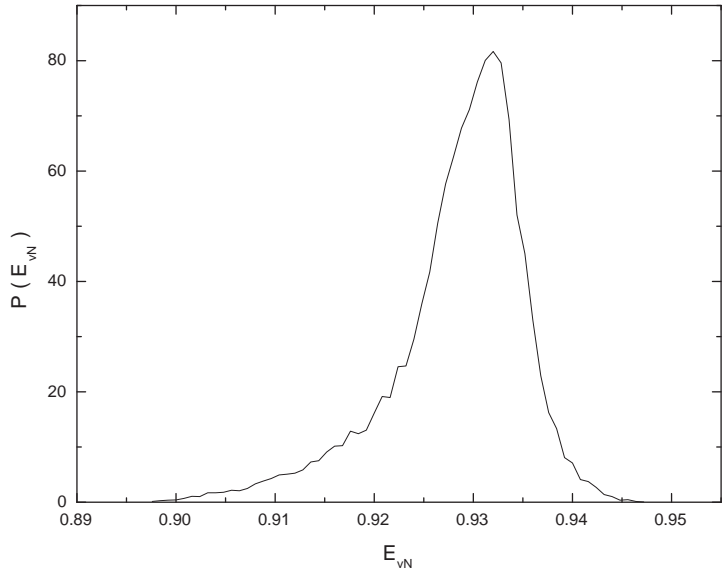}
\caption{Probability Density Function for
$E_{vN}$ among those states that maximize $E_L$.}
\end{center}
\label{Fig1}
\end{figure}

\noindent
with $\omega= -\frac{1}{2}+\frac{\sqrt{3}}{2}$. Although
it still remains unproven, several analytical
\cite{HS00,BH07} and numerical \cite{BSSB05,BPBZCP07}
evidences support the aforementioned conjecture.
In the cases of 5 and 6 qubits, states have been identified
having all their reduced density matrices maximally mixed
\cite{BSSB05,BPBZCP07}. Finally, for 7 qubits there is
numerical evidence suggesting that a recently discovered
state is the one with maximal entanglement although, as
in the 4 qubit case, no state of 7 qubits with all its
reduced density matrices maximally mixed was found
\cite{BPBZCP07}.

\section{Behaviour of $E_L$ and $E_{VN}$ for highly entangled states of 4 qubits.}

\begin{figure}
\begin{center}
\vspace{0.5cm}
\includegraphics[scale=0.4,angle=270]{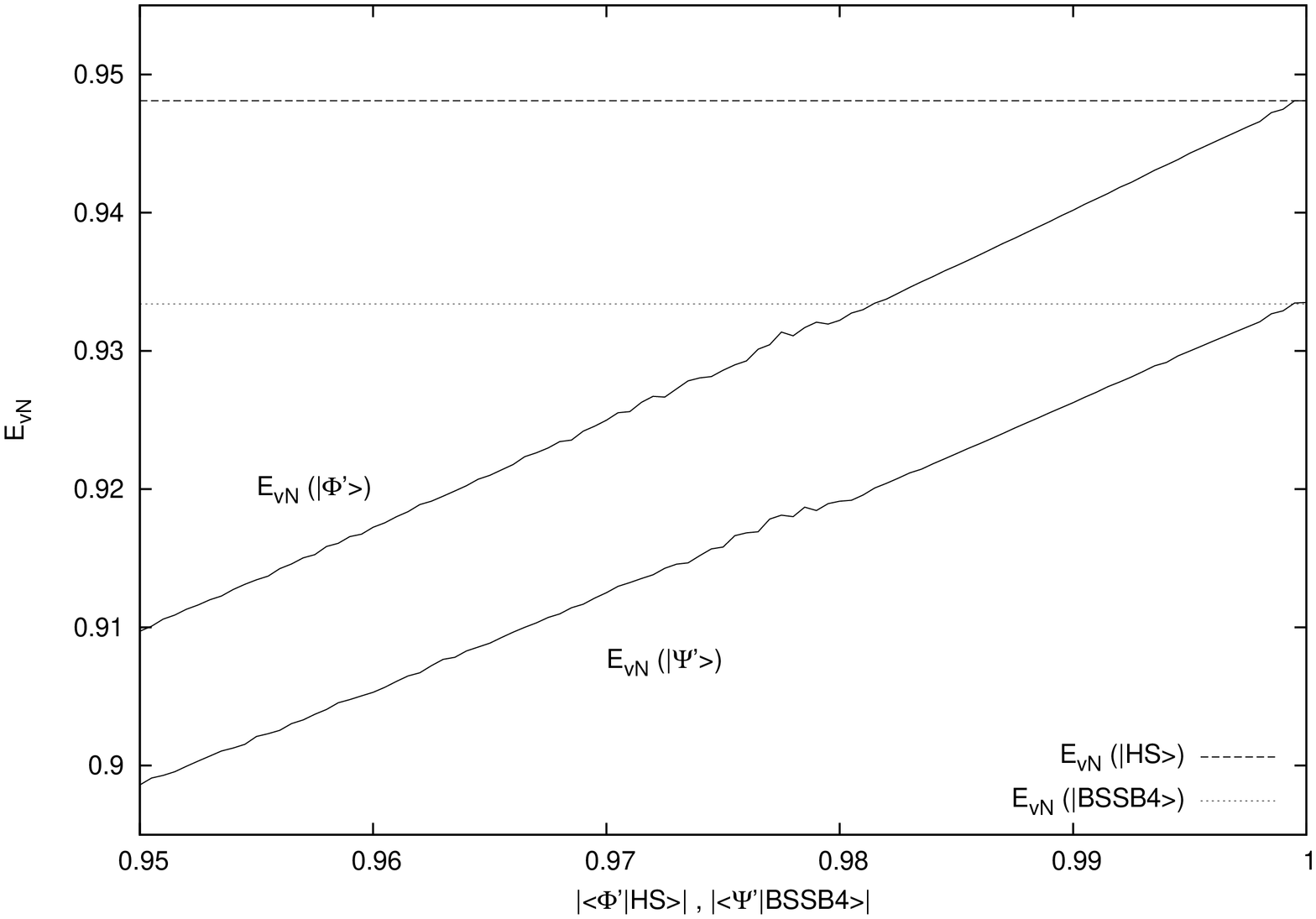}
\vspace {0.5cm}
\includegraphics[scale=0.4,angle=270]{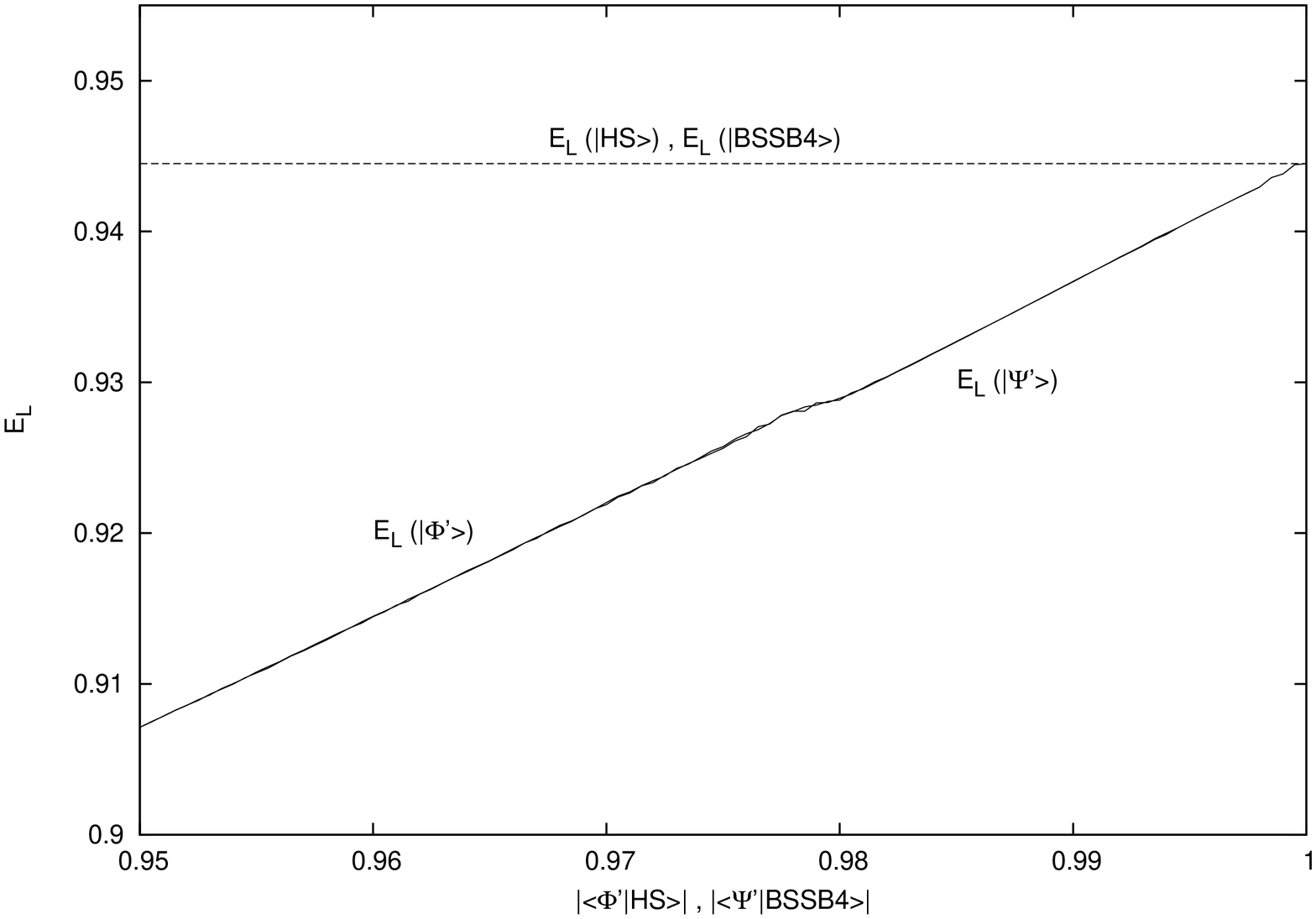}
\caption{Entanglement values for those states in the neighbourhood
of $|HS\rangle$ and $|BSSB4 \rangle$ as a function of the overlap with them.}
\end{center}
\label{Fig2}
\end{figure}

The multipartite entanglement measures $E_L$ based on the averaged
linear entropies of the reduced density matrices are widely used,
but sometimes it is more convenient to use an entanglement measure $E_{vN}$
based on the von Neumann entropy of the reduced density matrices,
although its computation is not as straightforward as the computation of
$E_L$. Here we compare the behaviour of the entanglement measures
$E_L$ and $E_{vN}$ when searching highly entangled states of 4 qubits.
Our results indicate that $E_{vN}$
is the best measure to use. As shown in Fig. 1, most states that
maximize $E_L$ are not maximally entangled according $E_{vN}$,
even though they are all highly entangled states (most of these
states have a value of $E_{vN}$ around $0.935$).

The study of the set of highly entangled 4 qubits states is of
considerable interest because they represent the lowest dimensional system
for which the non existence of the theoretically maximally entangled
state has been proved. Brown {\it et al}. \cite{BSSB05} developed
a numerical algorithm to search highly entangled states of
multi-qubit systems and found a maximally entangled state
of 5 qubits.  However, when applied to 4 qubit systems their
algorithm yielded a state (which we here call $|BSSB4 \rangle$)
less entangled than the $|HS \rangle$ state previously
discovered by Higuchi and Sudbery. This state is
$
|BSSB4 \rangle = \frac{1}{2} ( |0000\rangle +
|+011\rangle + |1101\rangle + |-110\rangle ),
$
where $|+\rangle = \frac{1}{\sqrt{2}}
( |0\rangle + |1\rangle )$ and
$|-\rangle = \frac{1}{\sqrt{2}} ( |0\rangle - |1\rangle )$.

 A new and slightly different numerical algorithm was recently
 developed by us \cite{BPBZCP07} that has been successfully applied
 to find maximally entangled states in systems up to 7 qubits,
including the 4 qubit $|HS \rangle$ state. Here we compare
the behaviour of $E_L$ and $E_{vN}$ as multiqubit entanglement
measures for highly entangled states of 4 qubits, through the
study of the entanglement properties of the states living in
the neighbourhoods of $|BSSB4 \rangle$ and $|HS \rangle$. To such an end we
first compute the average entanglement of states having given
overlaps with $|BSSB4 \rangle$. We considered, in total, a
family of $15 \, 000 \, 000$ states
$|\Psi ' \rangle$, with $0.95 \le |\langle \Psi ' |BSSB4 \rangle| \le 1$.
A similar computation is done with a second family of
 $15 \, 000 \, 000$ states $|\Phi ' \rangle$ (this time close to
 $|HS \rangle$) with $0.95 \le |\langle \Phi ' |HS \rangle| \le 1$.
The results are summarized in Fig. \ref{Fig2}. While both
entanglement measures identify all the alluded states
as highly entangled, $E_L$ does not succeed in distinguishing
the neighbours of $|BSSB4 \rangle$ from the neighbours of
$|HS \rangle$. On the other hand, the averaged $E_{vN}$ measure
successfully distinguishes both families of states, and identifies
the states in the neighbourhood of $|HS \rangle$ as more entangled than
those related to the $|BSSB4 \rangle$. Interestingly,
the slopes of both curves depicted in the upper part of Fig.2
(indicating the rate of decrease in entanglement
as we consider states with decreasing overlaps
with $|BSSB4 \rangle$ or $|HS \rangle$) are approximately
the same. This suggests that the ``entanglement landscapes" in the
neighbourhoods of $|BSSB4 \rangle$ or $|HS \rangle$
share some basic features.

\section{Alternative approach to the numerical search algorithm}

In Ref. \cite{BPBZCP07} we proposed a numerical search
algorithm that was able to find maximally entangled states
in systems up to 7 qubits, starting from an initial separable state.
To find the maximally entangled state, the coefficients of the
initial state are slightly modified to obtain a new one.
The entanglement of the new state is computed, if it is larger
than the entanglement of the previous state the new state is kept.
Otherwise, the new state is rejected and a new, tentative state
is generated. This iterative process is repeated until it converges
to a maximally entangled state.
At each iteration the entanglement given by Eq. (1) must be computed.
Consequently, at each step we must evaluate as many $E^{(m)}$
 measures as non-equivalent bi-partitions the system has.
This implies an exponential increase with the number of qubits of the
computational resources needed to find the final state.
Consequently, it is highly desirable to develop schemes to
decrease the number of iterations needed to obtain the convergence
and the time needed to perform each iteration.

\begin{table}[ph]
{\begin{tabular}{@{}cccccc@{}} \toprule
Number of qubits & 3 & 4 & 5 & 6 & 7  \\  \colrule
$E_L$ & 1.0000 & 0.9445 & 1.0000 & 1.0000 & 0.9961 \\
$E_{vN}$ & 1.0000 & 0.9481 & 1.0000 & 1.0000 & 0.9948 \\ \botrule
\end{tabular}}
\end{table}

If a multiqubit state has highly mixed reduced density matrices
corresponding
to subsystems of $[N/2]$ qubits, it is reasonable to
expect that the same will happen with the reduced
density matrices describing smaller subsystems. For this
reason, in order to optimize our algorithm, we have
tried a modified scheme based on the maximization of $E_{vN}^{[N/2]}$.
 The results of this experiment have been reasonably successful.
 For systems of 3, 4, 5, and 6 qubits the final highly
entangled states obtained maximizing $E_{vN}$ are the same as those obtained
maximizing $E_{vN}^{[N/2]}$. This is a big improvement in our numerical algorithm,
because in each iteration the number of bi-partitions to be considered
is roughly reduced to the half, and the total number of iterations
needed to reach the convergence are usually considerably reduced as
well. For 7 qubits the $E_{vN}$ entanglement
values of the states yielded by the $E_{vN}^{[N/2]}$-based algorithm
differ in the sixth decimal digit from
the $E$ entanglement value of the optimum state obtained maximizing $E_{vN}$.
   For 8 qubits the optimization algorithms, based
  either on balanced bi-partitions or on
the global entanglement measure, do not converge
always to the same state.
The entanglement values (for different number of qubits) of the
multiqubit states of highest entanglement considered in the
present work are given in Table 1.

\section{Conclusions} We have compared the behaviours of the multiqubit
entanglement measures $E_L$ and $E_{vN}$ based, respectively, on the
linear and the von Neumann entropies. Our results indicate that $E_{vN}$
is better than $E_L$ for the search of highly entangled states, because it
discriminates between states that, while exhibiting the same value of $E_L$,
have different degrees of entanglement. We also found evidence that
search algorithms based upon balanced
bi-partitions are almost as efficient as those based on the
complete set of bi-partitions.

\section*{Acknowledgments}

This work was partially supported by the MEC grant FIS2005-02796(Spain)
and FEDER(EU) and by CONICET (Argentine Agency).
AB acknowledges support from the FPU grant AP-2004-2962 (MEC-Spain).


\end{document}